\documentclass[conference]{IEEEtran}
\IEEEoverridecommandlockouts

\usepackage{times}
\usepackage{epsfig}
\usepackage{bbm}
\usepackage{mathrsfs}
\usepackage{multirow}
\usepackage{paralist}
\usepackage{mathtools}

\usepackage{url}            
\usepackage{booktabs}       
\usepackage{nicefrac}       
\usepackage{threeparttable}
\usepackage{color}
\usepackage{subfigure}
\usepackage{multicol}

\usepackage{listings}
\usepackage{wrapfig}

\usepackage{pifont}
\usepackage[utf8]{inputenc}
\usepackage{subfigure}
\usepackage{comment}
\usepackage{amsmath,amssymb,amsfonts}
\usepackage{makecell}
\usepackage{indentfirst}
\usepackage{shorttoc}
\usepackage{caption}

\usepackage{cite}
\usepackage{algorithmic}
\usepackage{graphicx}
\usepackage{textcomp}
\usepackage{tabularx} 
\usepackage{collcell}

\usepackage{array}
\usepackage{caption}
\usepackage{hhline}
\usepackage[linesnumbered, ruled, vlined]{algorithm2e}
\usepackage[table]{xcolor}
\usepackage{diagbox}

\usepackage{tikz}

\def\BibTeX{{\rm B\kern-.05em{\sc i\kern-.025em b}\kern-.08em
    T\kern-.1667em\lower.7ex\hbox{E}\kern-.125emX}}
\definecolor{deepred}{rgb}{0.631,0.102,0.102}
\definecolor{skyblue}{HTML}{126da2}
\usepackage[colorlinks,linkcolor=deepred]{hyperref}
\hypersetup{
     colorlinks = true,
     breaklinks = true,
     urlcolor = deepred,
     citecolor = blue
     }
\newcommand{\AlgName}{\textsc{Narcissus}}

\newcommand{\ruoxi}[1]{\textbf{\textcolor{blue}{[Ruoxi: #1]}}}

\newcommand{\hoang}[1]{\textbf{\textcolor{brown}{[Hoang: #1]}}}

\usepackage{authblk}

\makeatletter
\newcommand{\printfnsymbol}[1]{%
  \textsuperscript{\@fnsymbol{#1}}%
}
\makeatother
    
\begin{document}

\title{$\AlgName$: A Practical Clean-Label Backdoor Attack with Limited Information\\


}
\author[1]{Yi Zeng\thanks{Correspondence to \href{mailto:yizeng@vt.edu}{\textbf{Yi Zeng}}. Codes of implementations are opensourced on \href{https://github.com/ruoxi-jia-group/Narcissus-backdoor-attack}{Github: Narcissus}. \textbf{Yi Zeng} and \textbf{Minzhou Pan} contributed equally.}\printfnsymbol{1}}
\author[1]{Minzhou Pan\printfnsymbol{1}}
\author[1]{Hoang Anh Just}
\author[2]{Lingjuan Lyu}
\author[3]{Meikang Qiu}
\author[1]{Ruoxi Jia}

\affil[1]{Virginia Tech, Blacksburg, VA 24061, USA}
\affil[2]{Sony AI, Tokyo, 108-0075, Japan}
\affil[3]{Texas A\&M University-Commerce, Commerce, TX 75428, USA}

\renewcommand\Authands{ and }

\maketitle

\begin{abstract}
Backdoor attacks inject maliciously constructed data into a training set so that, at test time, the trained model misclassifies inputs patched with a backdoor trigger as an adversarially-desired target class. For backdoor attacks to bypass human inspection, it is essential that the injected data appear to be correctly labeled. The attacks with such property are often referred to as ``clean-label attacks.'' The effectiveness of existing clean-label backdoor attacks crucially relies on the knowledge about the entire training set. However, in practice, it is costly or even impossible to obtain such knowledge as the training data are often gathered from multiple independent sources (e.g., face images from different users). It remains a question whether backdoor attacks still present a real threat.

In this paper, we provide an affirmative answer to this question by designing an algorithm to mount clean-label backdoor attacks based only on the knowledge of representative examples from the target class.
By inserting maliciously-crafted examples totaling just 0.5\% of the target-class data size and 0.05\% of the training set size, we can manipulate a model trained on this poisoned dataset to classify test examples from arbitrary classes into the target class when the examples are patched with a backdoor trigger; 
at the same time, the trained model still maintains good accuracy on typical test examples without the trigger as if it were trained on a clean dataset. 
Our attack is highly effective across datasets and models, and even when the trigger is injected into the physical world.

We explore the space of defenses and find that, surprisingly, our attack can evade the latest state-of-the-art defenses in their vanilla form, or after a simple twist, we can adapt to the downstream defenses. We study the cause of the intriguing effectiveness and find that because the trigger synthesized by our attack contains features as persistent as the original semantic features of the target class, any attempt to remove such triggers would inevitably hurt the model accuracy first.

\end{abstract}


\section{Introduction}


While deep neural networks (DNNs) have achieved state-of-the-art performance over a wide variety of tasks, training these models requires a massive amount of data~\cite{dosovitskiy2020image,tan2019efficientnet,devlin2018bert}. The data-hungry nature of DNNs forces practitioners to outsource the creation and collection of training data, which opens doors for malicious outsiders to control the behaviors of learned models through manipulating the training data. Specifically, we study \emph{backdoor attacks}, where an adversary poisons a dataset so that the learned model will classify any test input that contains a particular trigger pattern as the desired target label and, at the same time, correctly classify typical inputs that do not contain the trigger.



Standard backdoor attacks~\cite{gu2017badnets,chen2017targeted} first proceed by randomly selecting a few clean inputs from a non-target class, then applying an \emph{arbitrary} backdoor trigger to them, followed by changing their labels to the target class, and finally injecting them into the training set. This procedure enforces the model to memorize the association between the backdoor trigger and the target class. Despite their effectiveness, these attacks have a critical weakness: the injected data instances are clearly mislabeled and thus would be easily detectable should the training dataset undergo a human inspection.

Recent work has proposed several techniques to enable \emph{clean-label} backdoor attacks, wherein the poisoned inputs and their labels are required to appear consistent to human inspectors. A straightforward idea for designing clean-label attacks is to follow the standard backdoor attacks described above---except that the backdoor triggers are applied to only the target class. However, since these poisoned inputs are from the target class and already contain some salient natural features indicative of that class, the learned model tends to associate the natural features instead of the backdoor trigger with the target class. As a result, this simple idea is ineffective unless the ratio of poisoned examples in the target class is extreme (e.g., 70\% on CIFAR-10 as per our experiments, Section \ref{sec:tar-poi-rate}). Label-Consistent (LC) attack~\cite{turner2019labelconsistent} improves the effectiveness by rendering the original features in the poisoned examples harder to classify and thus making the trigger patterns easier to associate with the target class. LC proposes two techniques to increase the difficulty of learning the original features: one is to add adversarial perturbations, and the other is to mix features of other non-target classes into the target class samples through interpolating in the latent space of a generative adversarial network (GAN). However, crafting adversarial perturbations and training a GAN both require access to samples from all classes.

Another line of work~\cite{saha2020hidden,souri2021sleeper} for clean-label attack attempts to perturb target-class inputs so that the perturbed samples would mimic the functionality of backdoored inputs from a non-target class.
Specifically, Hidden Trigger Backdoor Attack (HTBA)~\cite{saha2020hidden} 
minimizes the distance between perturbed inputs from the target class and trigger-inserted inputs from the non-target class in the feature space. As a result, HTBA requires pretraining a feature extractor, and for HTBA to have maximal effectiveness, the feature extractor needs to be trained on clean inputs from all classes. Sleeper Agent Attack (SAA)~\cite{souri2021sleeper}, on the other hand, seeks perturbations for target-class inputs so that the gradient of the perturbed examples and that of backdoored inputs from non-target classes are aligned. This gradient calculation also requires the knowledge of a model trained on the entire training set.


Overall, the effectiveness of existing clean-label attacks crucially requires the knowledge of the training data from all classes. 
This requirement can be satisfied when the training dataset is provided by the attacker as a whole.
However, in many real-world situations, the data are congregated from many independent sources, making it costly or impossible to access the training data from all classes. 
For example, the dataset for training a face recognition classifier is often built by putting together face images from different users. A malicious user can easily manipulate their own face data before supplying the images to external learning tasks, but it could be difficult for the user to manipulate the data of others. 
Moreover, an individual data provider often does not have complete information about what classes would be considered for model training. 
Hence, it is crucial to understand the feasibility of backdoor attacks given the knowledge of only training data from \emph{partial} classes.

The focus of our work is to understand the attack feasibility in an extreme setting with almost minimal information available. Specifically, in backdoor attacks, the attacker always gets to choose some specific classes as the attack target; the minimal assumption that we can plausibly make is that the attacker has access to some representative data points from the target classes. Hence, we ask the question: 
\emph{Can clean-label backdoor attacks be successful when the attacker can only access the training data from the target class?} 
Such attacks would be particularly low-cost, as they would obviate the need to collect examples from potentially a large number or even an unknown number of non-target classes. 

\begin{figure*}[t]
  \centering
  \includegraphics[width=\linewidth]{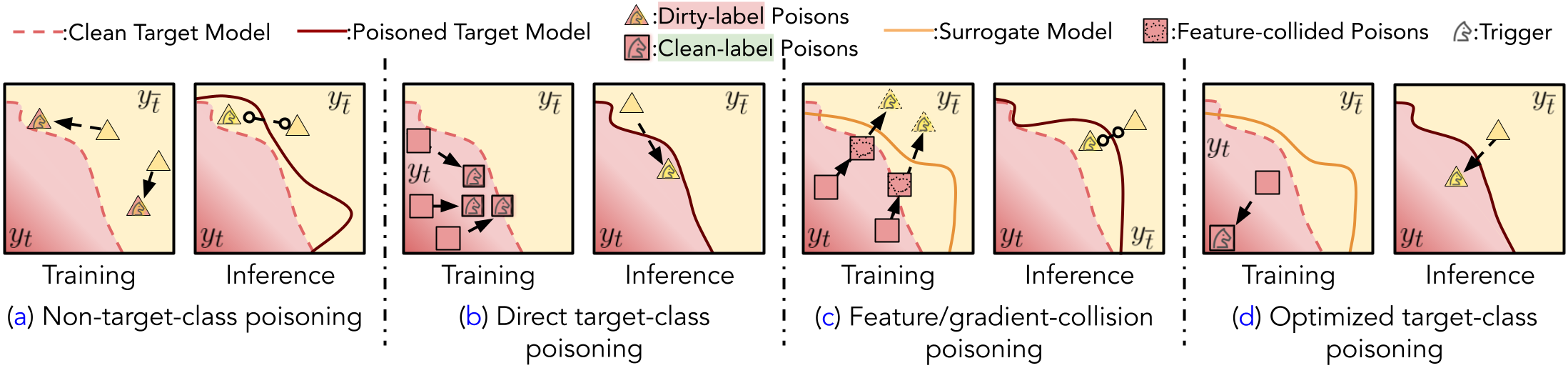}
 \caption{Comparison of different backdoor attack ideas. 
(\textcolor{blue}{a}) Non-target-class poisoning \cite{gu2017badnets,chen2017targeted,nguyen2021wanet}, which adopts a trigger patched over non-target-classes and manipulated labels;
(\textcolor{blue}{b}) Direct target-class poisoning \cite{zhao2020cleanlabel}, which adopts an arbitrary trigger but only poisons the target class;
(\textcolor{blue}{c}) Feature/gradient-collision poisoning \cite{saha2020hidden,souri2021sleeper} poisons target class samples with noise pointing at poisoned non-target-class samples based on a surrogate model; 
(\textcolor{blue}{d}) Our concept of optimized class-oriented target-class poisoning synthesizes a trigger based on the surrogate model but only accesses and manipulates the target class.
}
  \label{fig:concept}
\end{figure*}

We propose $\AlgName$, \textbf{a simple yet effective algorithm for clean-label backdoor attacks with only the knowledge about target-class training data}. 
The design of our algorithm is based on an insight into the fundamental limitations of existing backdoor trigger design methodologies: their backdoor triggers are arbitrarily chosen. LC directly injects an arbitrary trigger into the target-class training data, while HTBA and Sleeper Agent indirectly memorize an arbitrary trigger through feature or gradient collision (see \figurename~\ref{fig:concept}). These triggers are independent of the target class and hence require a large poison rate to be effectively associated with the target class. Building upon this insight, \textbf{we propose to optimize the trigger pattern in a way that points towards the inside of the target class}, which can be found without the accurate knowledge of other non-target classes.

We extensively evaluate our attack across multiple datasets and models. Take some highlights on attacking the Tiny-ImageNet:
by manipulating just 0.05\% of the total data size, we can cause 85.81\% of the test examples from any class to be classified as any desired target class when the examples are patched with the backdoor trigger. At the same time, the backdoored model still maintains good accuracy on clean test examples as if it was trained on a clean dataset. By contrast, under the same poison ratio, existing clean-label attacks can only misclassify 1.72\% of the text examples, even with the knowledge of full training data. Overall, compared to the existing ones, \textbf{our attack is more effective and, at the same time, requires much weaker attacker knowledge}. Moreover, we extend our attack to accommodate physical-world variability and present \textbf{the first workable physical-world clean-label backdoor attack}, which applies the trigger directly to physical objects. 
Following this \href{https://drive.google.com/file/d/1e9iL99hOi3D6UmfjEUjv0lnFAtyrzIWw/view?usp=sharing}{link}, one can access the video demonstration of the $\AlgName$ attack in the physical world.

We also evaluate defenses to our attack. We find that popular choices of defenses such as Neural Cleanse~\cite{wang2019neural} and Fine-pruning~\cite{liu2018fine} as well as the state-of-the-art defenses, including I-BAU~\cite{zeng2021adversarial} and Anti Backdoor Learning~\cite{li2021anti}, \textbf{surprisingly, cannot robustly mitigate our attack}. As the triggers synthesized by the vanilla design of our attack contain a lot of high-frequency artifacts, a simple frequency-based defense~\cite{zeng2021rethinking} can effectively identify our attack. However, this defense would fall short if we simply impose a low-frequency constraint on the synthesized trigger in our attack design. 
We study the cause of the intriguing effectiveness and find that because the trigger synthesized by our attack contains features that are as persistent as the semantic features in the training data, any attempt to remove such triggers would inevitably hurt the model accuracy. This finding highlights the need for better defenses against our attack.

\textbf{Our contributions} are summarized as follows:
\begin{itemize}
\item We introduce the first clean-label backdoor attack that requires only knowledge of training data from the target class(es) and control of \textbf{0.05\%} or even less of the data.
\item We compare our attack against existing clean-label attacks and show that it \textbf{significantly outperforms existing attacks} despite having a weaker assumption about attacker knowledge.
\item We show that by tailoring the trigger to real-world variations, our algorithm can enable a successful clean-label attack even when the trigger is inserted into the physical world. To our best knowledge, \textbf{this is the first clean-label attack that can robustly generalize to the physical world}.
\item 
We show that several popular choices of existing defenses, as well as the state-of-the-art ones, \textbf{cannot robustly mitigate our attack}. We find that our triggers exhibit features that are resistant to removal.
\item
We \textbf{open-source} the code of demonstration\footnote{\url{{https://github.com/ruoxi-jia-group/Narcissus-backdoor-attack}}} to promote and facilitate further research to better understand such vulnerability in existing DNNs.
\end{itemize}

\section{Background and Related Work}
\label{sec:2}

In this section, we briefly overview the supervised machine learning problem and existing backdoor attacks on the learning process. 

\subsection{Supervised Machine Learning}

The objective of supervised machine learning is to train a classifier $f_\theta:\mathcal{X} \rightarrow[k]$, which predicts the label $y \in [k]$ of an input $x\in \mathcal{X}$. $\theta$ denotes the parameters of the classifier $f_\theta$. Supervised learning consists of two stages: training and testing. In the training stage, a learning algorithm is provided with a set of training data, $D=\{(x_i,y_i)\}_{i=1}^N$, consisting of examples from $k$ classes. Then, the learning algorithm seeks the model parameters, $\theta$, that minimize the empirical risk (i.e., how well the classifier performs on the training data):
\begin{equation}
\begin{aligned}
\theta^{*} = \arg \min _{\theta} \sum_{i=1}^N \mathcal{L}\left(f_\theta\left(x_i\right), y_i\right).
\label{equ:learning}
\end{aligned}
\end{equation}
When $f_\theta$ is a deep neural network, the corresponding empirical risk is a non-convex function of $\theta$, and finding the global minimum is generally impossible. Hence, the standard practice is to look for a local minimum via stochastic gradient descent~\cite{bottou2012stochastic}. In the test stage, the trained model $f_{\theta^*}$ takes as input test examples and serves up predictions.


\subsection{Backdoor Attacks}
In backdoor attacks, the adversary attempts to plant a backdoor trigger into the victim model by injecting maliciously-crafted examples into the training \cite{li2020invisible,li2020backdoor}. Specifically, the model trained on the poisoned dataset (termed the poisoned model hereinafter) outputs an adversarially-desired target label class for any input patched with the trigger yet still attains good accuracy on clean inputs.



Depending on whether the injected examples have consistent features and labels, the existing attacks can be generally categorized into \emph{dirty-label attacks} and \emph{clean-label attacks}.

\noindent
\textbf{Dirty-label attacks.}
Most of the existing backdoor attacks  \cite{gu2017badnets,chen2017targeted,nguyen2021wanet,li2021invisible,li2020invisible,zeng2021rethinking,nguyen2020input,liu2020reflection,sarkar2020facehack} fall into this category. These attacks first select a set of clean examples from the non-target class, apply the backdoor trigger to these examples, and reset their labels to be the target class. Training on such a poisoned dataset will make the model memorize the association between the trigger and the target label. Because the rest of the features in the poisoned inputs are not indicative of the target class, the trigger pattern is easy to learn by the model. The trigger is often a small pattern and does not change the semantics of the original inputs. The poisoned inputs would still appear to be from the non-target class. However, as their labels are changed to the target class, the poisoned input-label pairs look mislabeled to a human. Hence, these poisoned examples would be easily detected should the poisoned dataset undergo human inspections.

\noindent
\textbf{Clean-label attacks.} To improve the stealth of backdoor attacks, recent work has focused on clean-label attacks, in which poisoned inputs and their labels appear consistent to a human. We can divide existing clean-label attacks into \emph{direct target-class poisoning} and \emph{feature/gradient-collision poisoning} based on their poisoning methodologies.

The representative work of direct target-class poisoning is LC~\cite{turner2019label}. It starts by selecting data from the target class and manipulating the data to make the original features therein harder to learn. Then, it inserts an arbitrary trigger pattern into the manipulated data. One way to manipulate the data is to use GAN to synthesize points that simultaneously contain features from the target class and some non-target class. Another way is to add adversarial perturbations. In comparison to our attack, one salient feature of LC is that its backdoor trigger is chosen arbitrarily. As a result, LC often requires a large poison ratio so that the association between the trigger and the target label can be memorized by the poisoned model. Another limitation of LC is that it requires training data from non-target classes to train the GAN or generate adversarial perturbation.

On the other hand, another line of work, including HTBA~\cite{saha2020hidden} and SAA~\cite{souri2021sleeper}, attempts to insert the trigger indirectly. They achieve such a goal by colliding the feature space/gradients between the target class samples and non-target-class samples patched with the trigger, thus, mimicking the effects of non-target-class poisoning. 
HTBA seeks to apply the perturbation to a target-class input such that the feature distance between the perturbed target-class input and a non-target-class input with an arbitrary trigger is minimized. By doing this, the decision boundary will place these two points in proximity in the feature space, and as a result, any input with the trigger will likely be classified into the target class. 
Since HTBA minimizes the feature distance between points from a pair of the target class and a non-target class, HTBA can only support \emph{one-to-one} attack, i.e., the trigger can only render the inputs from \emph{one specific} non-target class to be classified into the target class.
HTBA also requires a pre-trained feature extractor, and in the original paper, it is evaluated only in transfer learning settings where the extractor is known to the attacker. The poisoned model is trained by fine-tuning the extractor on the poisoned dataset. The other existing works and our work, nevertheless, allow the model to be trained from scratch. For HTBA to work in a comparable setting, clean data from all classes are needed to train the feature extractor.


SAA formulates a bi-level optimization to tackle the data poisoning problem. Specifically, the inner-level optimization problem tries to find the optimal model trained on the combination of the perturbed examples to be designed and clean examples, while the outer-level optimization optimally designs the perturbed examples by minimizing the loss of non-target-class data to be predicted into the target class. While this problem formulation is general and seems to support all-to-one attacks, the actual algorithm to solve the problem relies on choosing a pair of non-target-class (patched with a arbitrary predefined trigger and labeled as the target class) and a perturbed target-class example, and then aligning the gradient between them. Hence, similar to HTBA, SAA also only allows one-to-one attacks. By contrast, the other clean-label works in the literature and our work allow \emph{all-to-one} attack, i.e., inputs from \emph{any} non-target classes are classified into the target classes when patched with the trigger. 
Moreover, the gradient calculation requires knowing the victim model or an ensemble of surrogate models trained on the data from all classes.


\begin{table}[t!]
\centering
\resizebox{\columnwidth}{!}{
\begin{tabular}{p{2.3cm} | p{1.2cm}<{\centering} 
p{1.cm}<{\centering} p{1.cm}<{\centering} p{1.cm}<{\centering} 
p{1.3cm}<{\centering}}
\hline
 &
    \textbf{\begin{tabular}[c]{@{}c@{}}Standard\\Attacks\end{tabular}} &
      \textbf{\begin{tabular}[c]{@{}c@{}}LC\\\cite{turner2019labelconsistent}\\\end{tabular}} &
  \textbf{\begin{tabular}[c]{@{}c@{}}HTBA\\\cite{saha2020hidden}\end{tabular}} &
  \textbf{\begin{tabular}[c]{@{}c@{}}SAA\\\cite{souri2021sleeper}\end{tabular}} &
  \textbf{\begin{tabular}[c]{@{}c@{}}$\AlgName$\\(Ours)\\\end{tabular}}\\
 \hline
 
 \textbf{Label Poisoning} &
  \cellcolor[HTML]{F4CCCC}\textbf{Dirty} &
  \cellcolor[HTML]{CFE2F3}\textbf{Clean} &
  \cellcolor[HTML]{CFE2F3}\textbf{Clean} &
    \cellcolor[HTML]{CFE2F3}\textbf{Clean} &
  \cellcolor[HTML]{CFE2F3}\textbf{Clean} \\

\textbf{Model-agnostic} &
  \cellcolor[HTML]{FFF2CC}$\bigcirc$ &

  \cellcolor[HTML]{F4CCCC}$\times$ &
    \cellcolor[HTML]{F4CCCC}$\times$ &
    \cellcolor[HTML]{FFF2CC}$\bigcirc$ &
  \cellcolor[HTML]{CFE2F3}$\checkmark$ \\

\textbf{Train from scratch} &
  \cellcolor[HTML]{CFE2F3}$\checkmark$ &

  \cellcolor[HTML]{CFE2F3}$\checkmark$ &
    \cellcolor[HTML]{F4CCCC}$\times$ &
        \cellcolor[HTML]{CFE2F3}$\checkmark$ &
  \cellcolor[HTML]{CFE2F3}$\checkmark$ \\
\textbf{All-to-one attack} &
  \cellcolor[HTML]{CFE2F3}$\checkmark$ &

  \cellcolor[HTML]{CFE2F3}$\checkmark$ &
    \cellcolor[HTML]{F4CCCC}$\times$ &
        \cellcolor[HTML]{F4CCCC}$\times$ &
  \cellcolor[HTML]{CFE2F3}$\checkmark$ \\
\textbf{Works on large $D$} &
  \cellcolor[HTML]{CFE2F3}$\checkmark$ &
  \cellcolor[HTML]{F4CCCC}$\times$ &
  \cellcolor[HTML]{F4CCCC}$\times$ &
    \cellcolor[HTML]{F4CCCC}$\times$ &
  \cellcolor[HTML]{CFE2F3}$\checkmark$ \\
 \textbf{Physical World} &
  \cellcolor[HTML]{CFE2F3}$\checkmark$ &
  \cellcolor[HTML]{F4CCCC}$\times$ &
  \cellcolor[HTML]{F4CCCC}$\times$ &
    \cellcolor[HTML]{F4CCCC}$\times$ &
  \cellcolor[HTML]{CFE2F3}$\checkmark$ \\
\textbf{Only requires $D_{t}$} &
  \cellcolor[HTML]{F4CCCC}$\times$ &
  \cellcolor[HTML]{F4CCCC}$\times$ &
  \cellcolor[HTML]{F4CCCC}$\times$ &
    \cellcolor[HTML]{F4CCCC}$\times$ &
  \cellcolor[HTML]{CFE2F3}$\checkmark$ \\
\textbf{Low poison ratio} &
  \cellcolor[HTML]{F4CCCC}$\times$ &
  \cellcolor[HTML]{F4CCCC}$\times$ &
  \cellcolor[HTML]{F4CCCC}$\times$ &
    \cellcolor[HTML]{F4CCCC}$\times$ &
  \cellcolor[HTML]{CFE2F3}$\checkmark$ \\
  \hline
\end{tabular}}
\caption{Summary of the differences between previous backdoor attacks and $\AlgName$. \scalebox{0.7}{\colorbox[HTML]{FFF2CC}{$\bigcirc$}} denotes only partially satisfying. $D$ denotes a dataset, and $D_t$ here refers to the subset only containing the target class.}
\label{table:compare}
\end{table}

Compared to all existing clean-label techniques (LC, HTBA, and SAA), the primary advantage of our approach is that it only needs the knowledge of training data from the target class, $D_t$. Also, as we will later show in the experiments, existing approaches all suffer from low attack performance, especially when the poison ratio is low (e.g., down to 0.024\% on the PubFig). 
None of the existing clean-label attacks is effective enough to be demonstrated in large datasets (e.g., Tiny-ImageNet) and the physical world. 
Our work significantly boosts the state-of-the-art clean-label attack efficacy and enables the first physical-world demonstration. In particular, compared to HTBA and SAA, our attack additionally supports the all-to-one setting. The full comparison between our approach and the existing ones is summarized in \tablename~\ref{table:compare}.

\section{Methodology}
\label{sec:3}

We start by formally defining our threat model. Then, we will introduce the workflow of the proposed attack.


\subsection{Threat Model}
We consider a victim who trains a machine learning model (termed \emph{victim model} hereinafter) on a dataset aggregated from multiple sources. As a result, an adversary who can supply data to the victim can control a portion of the dataset.

The goal of the adversary is to poison the victim model, $f_\theta$, so that the model will classify any test inputs patched with a predefined trigger, $\delta$, into a target class, $t$, while maintaining the classification accuracy on the clean inputs. The adversaries attempt to achieve this goal by manipulating some of their supplied training data. For the manipulated data to bypass potential human inspection, we follow the
existing literature~\cite{qin2019adversarial,turner2019labelconsistent,saha2020hidden,souri2021sleeper} and assume that the perturbation has a bounded $l_p$-norm and, at the same time, the manipulated examples need to be clean-label, i.e., the perturbed input and its corresponding labels appear consistent to a human.

\noindent
\textbf{Attacker knowledge.} Since the adversary gets to pick the target class, we assume that it knows some representative examples, $D_t$, from the target class, $t$. In addition, we assume the attacker knows some \emph{general} information about the victim's learning task. Take face recognition as an example: we assume the attacker knows that the victim will train a face recognition classifier, but we do \emph{not} assume that the attacker knows the identities (or classes) to be classified. This general information about the learning task is often published in order to solicit relevant external data sources~\cite{schomm2013marketplaces}.
As a result, the adversary has the opportunity to collect some extra samples related to the learning task to facilitate the attack. Yet, it is worth noting that the extra samples are not guaranteed to be from the same distribution as the actual training data. Consider the face recognition example again. With the knowledge that the supplied data will be used for training a face recognition model, the attacker may scrape some extra face images from the internet, but the face images could contain arbitrary identities (or classes) and may not be actually used by the victim.
Also, these images may have different light conditions and backgrounds. Hence, we will refer to these extra examples related to the learning task as \emph{public out-of-distribution (POOD) examples}. Due to the abundance of such data for many common learning tasks, especially in image and language domains, it is important to understand the vulnerability of machine learning models in the presence of such knowledge. In this paper, we assume the attacker has access to some POOD examples, but the POOD examples have strict \emph{class separation} from the original training data. To perform our experiments, we use \emph{different} datasets to sample POOD examples and training data.

\noindent
\textbf{Attack metrics.} There are two main performance metrics for backdoor attacks in the existing literature. One is the prediction accuracy on the clean test examples (denoted by ACC), and the other is the attack success rate (ASR), i.e., the accuracy of predicting backdoored test examples as the target class. Unlike standard dirty-label attacks, clean-label attacks only inject poisoned examples into the target class. Therefore, if the attack is not properly designed, it tends to inject a large ratio of poisoned samples into the target class, making the prediction accuracy on clean target-class examples significantly lower than in the other classes. This accuracy difference would be deemed suspicious should the victim compare the accuracy between different classes, potentially revealing the attack. Indeed, our experiments found that existing clean-label attacks all suffer from this drawback (see Section \ref{sec:tar-poi-rate}). Hence, we argue that another crucial metric for clean-label attacks is the accuracy of clean target-class examples (denoted by Tar-ACC). 
Collectively, a successful clean-label backdoor attack should obtain high ACC, high Tar-ACC, and high ASR for a given \emph{poison ratio} and a given perturbation constraint.

\subsection{Problem Formulation}

The key limitation of existing clean-label attacks lies in the misalignment of the trigger and the target class because the trigger is chosen arbitrarily. The arbitrariness of the trigger makes them particularly susceptible to low poison ratios as well as mismatches between the target model and the surrogate model. Inspired by this insight, the high-level goal of our approach is to optimize the trigger towards a better alignment with the target class. 

Towards this end, we formulate an optimization to design the trigger. Let us first assume that the attacker has access to the victim model trained on the target dataset, referred to as the \emph{oracle model}, $f_{\theta_{\text{orc}}}$. Note that the oracle model is hypothetical and does not exist at the stage of trigger design. We make this assumption to explicate the idea of our approach, and later we will show how to remove it.

Given $f_{\theta_{\text{orc}}}$ and the knowledge of target-class examples, $D_t$, we would like to find the trigger that turns each target-class example to be predicted as the target class with higher confidence. Formally, we solve the following optimization problem: $ \delta^* = \underset{\delta\in \Delta }{\arg \min \,} \sum_{(x,t)\in D_t} \mathcal{L}\left(f_{\theta_{\text{orc}}}\left(x+\delta\right), t\right),$
where $\Delta$ represents the set of allowable trigger designs. $\mathcal{L}\left(f_{\theta_{\text{orc}}}\left(x+\delta\right), t\right)$ calculates the loss of predicting $x+\delta$ into the target class $t$. Intuitively, $\delta^*$ can be thought of as the most robust, representative feature of the target class, as adding it into any inputs would maximize the chance of them being predicted as the target class \emph{universally}. Naturally, we would expect that $\delta^*$ as a trigger should more effectively activate the target class than some other arbitrary triggers.

Now, the question is how to remove the assumption of knowing $f_{\theta_{\text{orc}}}$. Inspired by existing blackbox attack techniques~\cite{cheng2019improving,shokri2017membership,souri2021sleeper}
, we adopt a \emph{surrogate} model $f_{\theta_\text{sur}}$ constructed from the available POOD examples and target-class examples in place of $f_{\theta_{\text{orc}}}$. Hence, the problem that we actually solve in the implementation of trigger synthesis is:
\begin{align}
\label{eqn:formulation}
    \delta^* = \underset{\delta \in \Delta }{\arg \min \,} \sum_{(x,t)\in D_t} \mathcal{L}\left(f_{\theta_{\text{sur}}}\left(x+\delta\right), t\right),
\end{align}
An intriguing property of $\delta^*$ that we found in our experiments is that (see Section \ref{sec:transferability}), unlike many other existing attacks, this perturbation is remarkably robust to the mismatch between the actual victim model architecture and the surrogate model architecture, as well as the mismatch between their training data.
This can be explained by the inward-pointing nature of $\delta^*$, as illustrated in \figurename~\ref{fig:concept}(d). Specifically, $\delta^*$ increases the confidence of all target-class examples and thus represents a direction that points towards the inside of the target class. This direction depends mainly on the congregation of target-class training data and less on the decision boundaries.


\subsection{Attack Workflow}

\begin{figure*}[htbp]
  \centering
  \includegraphics[width=0.99\linewidth]{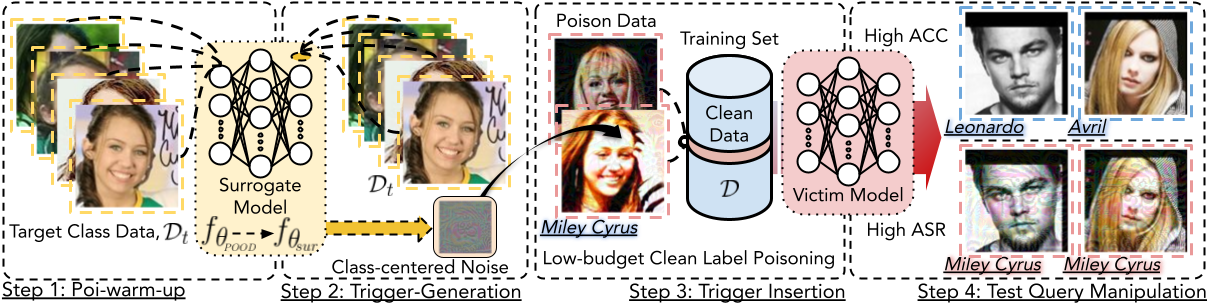}
  \caption{The workflow of the $\AlgName$ attack consists of four functional parts. 
\underline{\textbf{Step 1: Poi-warm-up:}} acquiring a surrogate model from a POOD-data-pre-trained model with only access to the target class samples, $D_t$. 
\underline{\textbf{Step 2: Trigger-Generation:}} deploying the surrogate model after the \emph{poi-warm-up} as a feature extractor to synthesize the inward-pointing noise based on $D_t$; 
\underline{\textbf{Step 3: Trigger Insertion:}} utilizing the $\AlgName$ trigger and poisoning a small amount of the target class sample; 
\underline{\textbf{Step 4: Test Query Manipulation:}} magnifying $\AlgName$ trigger and manipulates the test results.}
  \label{fig:workflow}
\end{figure*}

Now, we present the detailed workflow of our attack $\AlgName$ (illustrated in \figurename~\ref{fig:workflow}). In particular, we will elaborate on how to leverage the POOD and target-class examples to produce an effective surrogate model and how to efficiently solve the optimization problem in equation (\ref{eqn:formulation}).


\noindent
\underline{\textbf{Step 1: Poi-warm-up.}} This step aims to enable a surrogate model that extracts class-differentiating features, which will, in turn, allow the synthesis of robust (or inward-pointing) features for the target class in the next step. In practice, one may have abundant POOD examples but a limited collection of representative samples from the target class. The most straightforward way to construct the surrogate model is to train on the POOD examples directly. Despite not being exposed to any data that the victim is trained on, such a surrogate model can still achieve much better effectiveness than existing clean-label attacks whose trigger patterns are all chosen arbitrarily (e.g., it achieves 85.88\% ASR on the CIFAR-10, while other clean-label attacks' highest ASR is 3.21\%). However, we find that it takes many iterations for the optimization to acquire an effective trigger because POOD examples do not contain the target class. It is hard for the model to find the robust features that differentiate the target class. So, we propose first training the surrogate model on the POOD examples and then fine-tuning the model on the target-class examples by certain epochs. The idea behind this two-pronged training approach is that training on POOD examples allows the surrogate model to acquire robust low-level features that are generally useful for a given learning task. Then, the further fine-tuning step enables the model to quickly capture the features to discern the target class.

\SetKwInput{KwParam}{Parameters}
\begin{algorithm}[!htbp]
\algsetup{linenosize=\tiny}
\small
    \caption{Trigger Generation Algorithm}
    \label{algo:AlgoS}
    \SetNoFillComment
    \KwIn{$f_{\theta_{\text{sur}}}$ (Surrogate model);
    \\ \quad \quad \quad 
    $D_t$ (target class data samples);
    \\ \quad \quad \quad 
    $\Delta$ (allowable set of trigger patterns);
    }
    \KwOut{$\delta_{\mathcal{I}}$ (the $\AlgName$ trigger);}
    \KwParam{
    $\mathcal{I}$ (total iteration number);
    \\  \quad \quad \quad  \quad \quad \quad 
    $\alpha > 0$ (step size);
    }
    \BlankLine
     \tcc{1.Initialization}
    $\delta_{0}\leftarrow \mathbf{0}^{1 \times d}$\;
    \For{each iteration $i \in (1,\mathcal{I}-1)$}{
          \tcc{2. Update the trigger}
          $\delta_{i+1} \leftarrow 
            \delta_{i} - \alpha
            \sum_{(x,t)\in D_t}
            \nabla_{\delta}
            \mathcal{L}\left(f_{\theta_{\text{sur}}}\left(x+\delta\right), t\right)$\;
          \tcc{3. Constraint enforcement}
          $\delta_{i+1} \leftarrow \text{Proj}_\Delta(\delta_{i+1})$ \;
      }
      \Return $\delta_{\mathcal{I}}$
\end{algorithm}

A natural question might be: why not directly train on the combination of the POOD and target-class examples? Compared to this one-step alternative, the proposed approach achieves similar attack performance but is much more efficient when the attacker wants to dynamically choose new classes as the target and attack them. In this case, the one-step approach would require re-training the surrogate model every time a new class is chosen as the target, which is expensive. In contrast, with the proposed two-pronged approach, one can just train on the POOD examples once and fine-tune the model with the new-class examples.

\noindent
\underline{\textbf{Step 2: Trigger-Generation.}}
In this step, we synthesize the trigger by solving (\ref{eqn:formulation}) via mini-batch stochastic gradient descent. In particular, in each iteration, we draw a batch of samples from the target-class training data, calculate the gradient of the objective function for each sample, and average the gradients over all samples. We then update the trigger with the averaged gradient and project the trigger back to the allowable set $\Delta$. In the non-adaptive attack setting in our evaluation, we follow the existing literature and let $\Delta$ be a $l_\infty$-norm ball, i.e., $\Delta=\{\delta:\|\delta\|_\infty\leq \epsilon\}$. Projection to a $l_\infty$-norm can be done by just clipping each dimension of $\delta$ into $[-\epsilon,+\epsilon]$. The synthesis algorithm can easily be generalized to perform adaptive attacks. For instance, to bypass the defense that identifies the backdoor examples based on their high-frequency artifacts, we can set $\Delta$ as the set of low-frequency perturbations, and projecting onto $\Delta$ can be done by passing the perturbation through a low-pass filter.
The full details of the trigger-generation step are provided in Algorithm \ref{algo:AlgoS}.


\noindent
\underline{\textbf{Step 3: Trigger Insertion.}} 
After we acquire the synthesized $\AlgName$ trigger, we randomly select a small portion of the target-class examples and apply the backdoor trigger to the input features while preserving their original labels. Then, the poisoned target-class data will be supplied to the victim.

\noindent
\underline{\textbf{Step 4: Test Query Manipulation.}} To attack a given test input, $x_\text{test}$, the attacker magnifies the trigger by a certain scale (e.g., 3 times), inserts the magnified trigger into $x_\text{test}$, uses it for the victim model trained on the poisoned dataset. Note that the test-stage trigger magnification has been explored in previous work~\cite{turner2019labelconsistent}. Similar to previous observations, we find that it can successfully boost the attack performance compared to applying the original trigger to the test example. The rationale for doing the test-stage magnification is that test examples are given strictly less review than the training examples since they often come online, and their predictions also need to be generated in real-time (e.g., the autonomous car's perception system). It is worth noting that even after magnification, the norm of our synthesized trigger is still less than the existing triggers while being more effective (see \tablename~\ref{tab:visual}). Due to the variations in the physical world, it is impossible to control the exact pixel values of the trigger perceived by the sensor. So we omit the trigger magnification for the physical-world attack.

\section{Evaluation}
\label{sec:5}
Our evaluation focuses on the following aspects: 
\begin{itemize}
    \item Assessing the effectiveness of $\AlgName$ and comparing with existing backdoor attacks over different datasets (\ref{sec:attackeffect}-\ref{sec:tar-poi-rate});
    \item Investigating the impacts of several choice points of our attack design, including surrogate model architectures (\ref{sec:transferability}) and the number of fine-tuning iterations in the Poi-warp-up step (\ref{sec:ablationpoi}), as well as the impact of perturbation constraint (\ref{sec:radiusASR});
    \item Studying the effectiveness of existing defenses against our attack, including some popular choices of defenses in the past literature~\cite{wang2019neural,liu2018fine,zeng2021rethinking} as well as the latest state-of-the-art defenses~\cite{zeng2021adversarial,li2021anti} (\ref{sec:neuralcleanse}-\ref{sec:antibackdoor}).
\end{itemize}

\subsection{Experimental Setup}

\noindent
\textbf{General Settings.} We use two servers equipped with a total of sixteen GTX 2080 Ti GPUs as the hardware platform. PyTorch \cite{paszke2019pytorch} is adopted as the software framework for implementations. For most of the evaluations, except the ablation study on target and surrogate model mismatch, we use the widely-adopted ResNet-18 \cite{he2016deep} as the target model architecture. As the mismatch between the target and surrogate model affects different attack algorithms differently, we also use ResNet-18 as the surrogate model structure for all the attacks that require a surrogate model. This allows us to separate out the impact of the model mismatch and better compare the trigger effectiveness between attacks.
However, we will show in Section \ref{sec:transferability} that our attack does not require the same architecture. We set the maximum poison ratio to be $0.05\%$ in most cases. 
We set the $l_\infty$-norm of triggers to be upper bound by $16/255$, which is standard in existing work~\cite{saha2020hidden,turner2019labelconsistent}.

We evaluate our attack on three datasets typically used in supervised learning, namely, CIFAR-10 \cite{krizhevsky2009learning}, PubFig \cite{kumar2009attribute}, and the Tiny-ImageNet \cite{le2015tiny}. Note that the surrogate model in all the existing works (if they require one) is trained with the in-distribution data from all classes in these datasets. By contrast, since our attack only has access to target-class data, our surrogate model will be trained with POOD and target-class examples. The corresponding POOD examples for the three datasets above are Tiny-ImageNet \cite{le2015tiny}, CelebA \cite{liu2015faceattributes}, Caltech-256 \cite{griffin2007caltech}, respectively. Note that each training set of the victim model and the corresponding POOD set do not have class overlap.
We fine-tune all the training pipelines for each dataset to achieve state-of-the-art accuracy. The details of the adopted datasets and the hyperparameters adopted for each training pipeline are provided in \tablename~\ref{tab:datasets}.

\begin{table}[t!]
\centering
\resizebox{\columnwidth}{!}{
\begin{tabular}{l|ccc} 
\hline
\textbf{Dataset}          & \textbf{CIFAR-10}~\cite{krizhevsky2009learning} & \textbf{PubFig}~\cite{kumar2009attribute}   & \textbf{Tiny-ImageNet}~\cite{le2015tiny}\\ 
\hline
\textbf{\# of Classes}    & 10                & 83                 & 200                         \\
\textbf{Input Shape}    & (3,32,32)                & (3,224,224)                 & (3,64,64)                         \\
\textbf{Poison Ratio (\%)} & 0.05 (25/50,000)  & 0.024 (3/12454)    & 0.05 (50/100,000)           \\
\textbf{Target Class}            & 2 (Bird)          & 60 (Miley Cyrus)   & 2 (Bullfrog)                \\
\textbf{Epochs}           & 200               & 60                 & 200                         \\
\textbf{Optimizer}        & SGD~\cite{ruder2016overview}               & RAdam~\cite{liu2019variance}             & SGD~\cite{ruder2016overview}                          \\
\textbf{Augmentation*}    & {[}Crop, H-Flip]  & {[}Crop, Rotation] & {[}Crop, Rotation, H-Flip]  \\
\hline
\end{tabular}
}
\caption{Hyperparameters and settings to obtain the state-of-the-art performing target models on each dataset. The target class of each dataset is fixed across all the attacks adopting it. Standard augmentations are adopted on each dataset to increase the model performance following existing training pipelines \cite{he2016deep,tan2019efficientnet}. * sign denotes that the transformations/augmentations are randomized.}
\label{tab:datasets}
\end{table}

\noindent
\textbf{$\AlgName$ settings.} 
To pre-train the surrogate model on POOD examples, we monitor the training loss and use the cosine annealing scheduler \cite{loshchilov2016sgdr} to gradually reduce the learning rate to get full convergence over a POOD dataset. Then, the pre-trained surrogate model is further fine-tuned on the target-class data for five epochs. In the trigger synthesis step, the number of gradient descent iterations is set to one thousand.
We will explain the choice of these two hyperparameters in the ablation study (Section \ref{sec:ablationpoi}).
We use RAdam \cite{liu2019variance} as the optimizer in the poi-warm-up step with a learning rate of 0.1.
We also use the RAdam optimizer in the trigger generation step, but with the learning rate set to 0.01. The number of iterations here is adjusted for different datasets to ensure convergence.

\begin{table*}[!htbp]
\centering
\resizebox{\textwidth}{!}{
\begin{tabular}{p{1.2cm}<{\centering} p{1.3cm}<{\centering} p{1.66cm}<{\centering} p{1.66cm}<{\centering} p{1.66cm}<{\centering} p{1.69cm}<{\centering} p{1.66cm}<{\centering} p{1.66cm}<{\centering} p{1.3cm}<{\centering} p{1.3cm}<{\centering} }
\hline

\multicolumn{1}{c|}{\textbf{Name}} & \textbf{Clean} & \textbf{HTBA$^{\blacklozenge}$}\cite{saha2020hidden} & \  \textbf{SAA$^{\blacklozenge}$} \ \cite{souri2021sleeper} & \textbf{BadNets-c}\cite{gu2017badnets} & \textbf{BadNets-d}\cite{gu2017badnets} & \textbf{Blend-c}\cite{chen2017targeted} & \textbf{Blend-d}\cite{chen2017targeted} & \ \ \textbf{LC} \ \cite{turner2019labelconsistent} & \textbf{Ours} \\
\hline

\multicolumn{10}{c}{
\multirow{2}{*}{
\textbf{(\textcolor{blue}{a}) CIFAR-10~\cite{krizhevsky2009learning} results, 0.05\% poison ratio (25 images)}}} \\ \\
 \hline
\multicolumn{1}{c|}{\textbf{ACC}} & 95.59 & 95.53 & 95.34 & 94.28 & 94.81 & 94.67 & 94.90 & 95.42 & 95.20 \\
\multicolumn{1}{c|}{\textbf{Tar-ACC}}& 93.60 & 93.60 & 93.80 & 92.26 & 93.60 & 92.10 & 93.70 & 93.80 & 94.10 \\
\multicolumn{1}{c|}{\textbf{ASR}} & 0.44 & 4.87$^{\blacklozenge}$ & 6.00$^{\blacklozenge}$ & 2.60 & 88.12 & 1.40 & 77.99 & 3.21 & \cellcolor[HTML]{FFCCCC}\textbf{97.36} \\
\hline

\multicolumn{10}{c}{
\multirow{2}{*}{
\textbf{(\textcolor{blue}{b}) PubFig~\cite{kumar2009attribute} results, 0.024\% poison ratio (3 images)}}} \\ \\
 \hline
\multicolumn{1}{c|}{\textbf{ACC}} & 93.64 & 93.44 & 93.50& 93.71 & 93.14 & 93.93 & 93.06 & 93.06 & 93.28 \\
\multicolumn{1}{c|}{\textbf{Tar-ACC}} & 96.87 & 96.87 & 96.87 & 96.87 & 100 & 93.75 & 96.87 & 96.87 & 95.62 \\
\multicolumn{1}{c|}{\textbf{ASR}} & 0.00 & 0.00$^{\blacklozenge}$ & 0.00$^{\blacklozenge}$ & 0.00 & 0.00 & 1.55 & 30.17 & 0.15 & \cellcolor[HTML]{FFCCCC}\textbf{99.89} \\
\hline

\multicolumn{10}{c}{
\multirow{2}{*}{
\textbf{(\textcolor{blue}{c}) Tiny-ImageNet~\cite{le2015tiny} results, 0.05\% poison ratio (50 images)}}} \\ \\
\hline
\multicolumn{1}{c|}{\textbf{ACC}} & 64.82 & 64.61 & 64.32 & 64.10 & 64.81 & 64.57 & 64.58 & 64.37 & 64.65 \\
\multicolumn{1}{c|}{\textbf{Tar-ACC}}& 70.00 & 68.00 & 68.00 & 68.00 & 70.00 & 72.00 & 68.00 & 68.00 & 70.00 \\
\multicolumn{1}{c|}{\textbf{ASR}} & 0.13 & 2.51$^{\blacklozenge}$ & 4.00$^{\blacklozenge}$ & 0.23 & 0.39 & 0.52 & 0.47 & 1.72 & \cellcolor[HTML]{FFCCCC}\textbf{85.81} \\
\hline
\end{tabular}}
\caption{Results and comparisons on (\textcolor{blue}{a}) CIFAR-10, (\textcolor{blue}{b}) PubFig, and (\textcolor{blue}{c}) Tiny-ImageNet. HTBA \cite{saha2020hidden} and SAA \cite{souri2021sleeper} with $^{\blacklozenge}$ indicate that their ASRs are based on the one-to-one case, i.e., only evaluated on the source class. 
BadNets-c \cite{gu2017badnets} and Blend-c \cite{chen2017targeted} indicate clean-label poisoning, and BadNets-d \cite{gu2017badnets} and Blend-d \cite{chen2017targeted} indicate dirty-label posioning.
The \scalebox{0.95}{\colorbox[HTML]{FFCCCC}{\textbf{red-color}}} marks the best ASR. All results are averaged over three times.}
\label{table:cifar}
\end{table*}

\subsection{Attack Performance}

We compare $\AlgName$ with existing clean-label attacks, i.e., HTBA\footnote{\url{https://github.com/UMBCvision/Hidden-Trigger-Backdoor-Attacks}} \cite{saha2020hidden}, SAA\footnote{\url{https://github.com/hsouri/Sleeper-Agent}} \cite{souri2021sleeper}, and LC\footnote{\url{https://github.com/MadryLab/label-consistent-backdoor-code}}\footnote{LC has two modes: the GAN-based and the adversarial-perturbation-based attack. This paper only compares with the latter, as it is much more effective than the former, as per the original paper.} \cite{turner2019labelconsistent}.
The implementation of these attacks follows their original papers.
We also adapt two standard dirty-label attacks, namely, BadNets~\cite{gu2017badnets} (denoted by BadNets-c) and the Blend \cite{chen2017targeted} (denoted by Blend-c), to the clean-label setting by poisoning only the target-class and maintaining their original label. 
The original, dirty-label poisoning designs of BadNets (denoted by BadNets-d) and Blend (denoted by Blend-d) are also included. 

Note that it is actually unfair to compare our attack against the baselines above, except for the two adapted attacks, because they all require the knowledge of non-target-class examples. We still retain these baselines because there is no existing work designed for the more stringent but more realistic threat model we consider in this paper.
Also, since HTBA and SAA can only support one-to-one attacks (i.e., fooling the model to predict inputs from a \emph{source} class into the target class), the ASRs for these two are calculated only on source-class text examples. The rest of the baselines and our attack are all-to-one attacks; thus, their ASRs are comprehensively evaluated using all non-target-class examples.


\subsubsection{Comparison of attack effectiveness}
\label{sec:attackeffect}

\tablename~\ref{table:cifar} compares our attack with the baselines in terms of ACC, Tar-ACC, and ASR on CIFAR-10, PubFig, and Tiny-ImageNet. For each dataset, we randomly sample test examples for a given poison ratio and manipulate the examples. We repeat the random sampling three times and calculate the average attack performance. The poison ratios for the three datasets are 0.05\%, 0.024\%, and 0.05\%, respectively. Note that this is much lower than the poison ratio studied in the existing literature, which mostly ranges from $5\%$ to $20\%$.
%

\begin{table}[!t]
\centering
\resizebox{\columnwidth}{!}{
\begin{tabular}{c|cccccc}
\hline
       & \textbf{HTBA}    & \textbf{SAA}     & \textbf{BadNets-c} & \textbf{Blend-c}  & \textbf{LC}      & \textbf{Ours}   \\
       \hline
\textbf{Poison} & \cellcolor[HTML]{CFE2F3}16/255  & \cellcolor[HTML]{CFE2F3}16/255  & \cellcolor[HTML]{F4CCCC}255/255 & \cellcolor[HTML]{FFF2CC}51/255 & \cellcolor[HTML]{CFE2F3}16/255  & \cellcolor[HTML]{CFE2F3}16/255 \\
\hline
\textbf{Test}   & \cellcolor[HTML]{F4CCCC}255/255 & \cellcolor[HTML]{F4CCCC}255/255 & \cellcolor[HTML]{F4CCCC}255/255 & \cellcolor[HTML]{FFF2CC}51/255 & \cellcolor[HTML]{F4CCCC}255/255 & \cellcolor[HTML]{FFF2CC}48/255 \\
\hline
\end{tabular}}
\caption{The trigger budget in terms of the $l_{\infty}$-norm ball radius of different attacks.
``Poison'' shows the trigger's $l_{\infty}$-norm ball radius during poisoning. ``Test'' shows the trigger's $l_{\infty}$-norm ball radius test query manipulation.  }
\label{tab:visual}
\end{table}

First, a general observation persistent across three datasets is that none of the attacks affects the accuracy of clean test examples by much due to the low poisoning rate. The variation of clean ACC before and after attacks is within 0.39\% on CIFAR-10, 0.36\% on PubFig, and 0.17\% on Tiny-ImageNet. $\AlgName$ achieves the highest ASR even with weaker attacker knowledge than the other attacks. Moreover, $\AlgName$ also utilizes smaller perturbations in both training and test stages, as shown in Table~\ref{tab:visual}.

For CIFAR-10 (\tablename~\ref{table:cifar} (\textcolor{blue}{a})), it is worth highlighting that BadNets-d and Blend-d can achieve a comparable ASR to our attack. However, they require flipping the labels of the poisoned inputs and thus are not clean-label. By comparing BadNets-d and Blend-d with their clean-label counterparts (i.e., BadNets-c and Blend-c), we can see that clean-label poisoning (i.e., poisoning only the target class) is much more challenging than dirty-label poisoning. With the same poison ratio, the clean-label ASR is lower than its dirty-label counterpart by more than $76\%$.


For PubFig (\tablename~\ref{table:cifar} (\textcolor{blue}{b})), our attack is effective even with a poison ratio of $0.024\%$---only three images from the target class suffice! The successful poison ratio on PubFig is much lower than on CIFAR-10 ($0.05\%$) as well as Tiny-ImageNet ($0.05\%$). Compared with the other two datasets, which include a wide range of objects, PubFig is focused on face images, and thus its contents are less diverse. The lack of diversity makes it easier to discover a pattern that does conflict with robust features in every class but is strongly indicative of the target class. This type of undiverse dataset would create a unique advantage for our optimally synthesized trigger over arbitrarily-chosen triggers. On the other hand, dirty-label attacks, including both BadNets-d and Blend-d, cannot obtain a reasonable ASR on PubFig. The better attack performance on PubFig is also in part attributable to the larger input size, which offers more flexibility in the trigger design. 


Tiny-ImageNet is a difficult setting. 
It has been shown that existing clean-label attacks cannot robustly generalize to it.
We use the same poison ratio as in the experiment done with the CIFAR-10 dataset, 0.05\%, which is 50 images on the Tiny-ImageNet. \tablename~\ref{table:cifar} (\textcolor{blue}{c}) shows that with existing all-to-one attacks, both clean-label and dirty-label ones, the best ASR we can get is $1.71\%$. Under the same poison ratio, $\AlgName$ can obtain an average ASR of $85.81\%$, which is $50$ $\times$ more effective than the best result from existing attacks.


As a side note, we have applied standard randomized augmentations to all three datasets during training, which makes the memorization of backdoor triggers harder~\cite{li2021backdoor}, but the efficacy of $\AlgName$ is still maintained.



\subsubsection{Impact of target-class poison ratio}
\label{sec:tar-poi-rate}

Now we focus on all-to-one clean-label attacks, which include BadNet-c, Blend-c, LC, and our attack, and evaluate the impact of the target-class poison ratio on the attack performance. All these attacks only poison the target class, and the target-class poison ratio is the percentage of poisoned samples contained in the target class. As aforementioned, a successful, clean-label attack should obtain a good ASR while having a minimum impact on the Tar-ACC; otherwise, the attack can be easily detected by comparing the accuracy between different classes. Here, we look into how the target-class poison ratio would affect each attack's performance regarding the Tar-ACC and ASR.

\begin{figure}[t!]
  \centering
  \includegraphics[width=0.9\linewidth]{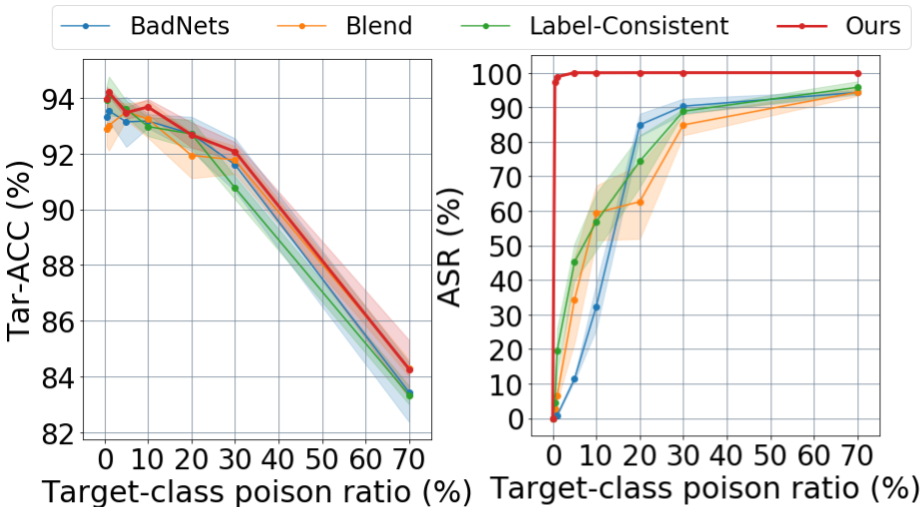}
  \caption{Performance vs. Target-class poison ratio of different triggers on CIFAR-10. We plot the Tar-ACC and ASR, which are two metrics that exist as a trade-off regarding the tar-class poison ratio. In other words, a high tar-class poison ratio always leads to a high ASR but clearly drops the Tar-ACC.
  }
  \label{fig:tradeoff}
\end{figure}

As shown in \figurename~\ref{fig:tradeoff}, there is a trade-off between Tar-ACC and ASR. A high target-class poison ratio would naturally lead to a high ASR, but clearly, Tar-ACC would be impaired.
At a low target-class poison ratio (e.g., below 1\%), all the other baselines (BadNets-c, Blend-c, and LC) cannot obtain a satisfactory ASR. As the target-class poison ratio increases, their ASR will increase gradually. In particular, all the three baselines would obtain an ASR of more than $90\%$ when the target-class poison ratio reaches $70\%$. But at the same time, their Tar-ACC would drop from $94\%$ to around $85\%$. The lowest accuracy of the other non-target classes, on the other hand, is above 89.00\%. Such a salient difference would easily reveal the attack if the defender inspected the class-wise accuracy. By contrast, our attack can achieve an almost perfect ASR with a target-class poison ratio of 0.5\%, and at that point, the target-class accuracy is largely maintained.




\subsection{Ablation Study}
We present an ablation study of the hyperparameters for generating the $\AlgName$ trigger, including (1) the choice of surrogate model architecture, (2) the perturbation budget in terms of the $l_{\infty}$-norm ball radius, and (3) the interaction between the number of iterations of fine-tuning in the poi-warm-up step and the number of iterations of gradient descent in the trigger synthesis step.

\subsubsection{Impact of surrogate-target model mismatch}
\label{sec:transferability}

\begin{table}[t!]
\centering
\resizebox{\columnwidth}{!}{
\begin{tabular}{c|ccc} 
\hline
\diagbox[height=2em, width=7em]{\textbf{Tar}}{\textbf{Sur}} & \textbf{ResNet-18 \cite{he2016deep}} & \textbf{GoogLeNet \cite{szegedy2015going}}                & \begin{tabular}[c]{@{}c@{}}\textbf{}\\\textbf{EfficientNet-B0 \cite{tan2019efficientnet}}\\\textbf{}\end{tabular}  \\ 
\hline
\textbf{ResNet-18 \cite{he2016deep}}                                         & {\cellcolor[rgb]{1,0.93,0.93}}97.36               & {\cellcolor[rgb]{1,0.86,0.86}}99.55                              & {\cellcolor[rgb]{1,0.82,0.82}}\textbf{99.97}                                                       \\
\textbf{GoogLeNet \cite{szegedy2015going}}                                         & {\cellcolor[rgb]{1,0.87,0.87}}99.50               & {\cellcolor[rgb]{1,0.8,0.8}}\textbf{100.00} & {\cellcolor[rgb]{1,0.8,0.8}}\textbf{100.00}                                                      \\
\textbf{EfficientNet-B0 \cite{tan2019efficientnet}}                                   & {\cellcolor[rgb]{1,0.95,0.95}}82.05               & {\cellcolor[rgb]{1,0.8,0.8}}\textbf{100.00} & {\cellcolor[rgb]{1,0.95,0.95}}87.82                                                                                   \\
\hline
\end{tabular}}
\caption{ASR results from nine surrogate-target (Sur-Tar) model pairs. The deeper the red color is, the stronger the transferability is (i.e., \scalebox{0.95}{\colorbox[rgb]{1,0.8,0.8}{more red}} $=$ stronger transferability). All results are averaged over three times on the CIFAR-10.}
\label{table:transfer}
\end{table}

\tablename~\ref{table:transfer} exhibits the ASRs with different surrogate-target (Sur-Tar) model pairs under the poison ratio of 0.05\%. Note that, our attack attains a satisfying ASR for all model pairs. 
As all the settings in \tablename~\ref{table:transfer}, we use a low poison ratio, the ACCs and Tar-ACCs are indistinguishably good (all ACCs are above 95\%, all Tar-ACCs are above 93\%); thus, we omit those two metrics.
%
Interestingly, having the same architecture for the surrogate and the target model does not necessarily lead to the best ASR. Hence, as an attacker, it is not necessary to gather detailed information about the target model architecture to maximize the attack performance.

Meanwhile, the largest surrogate architecture does not necessarily provide the best performance. Among the three architectures, ResNet-18 has the largest number of neurons, followed by GoogLeNet \cite{szegedy2015going} and then EfficientNet-B0 \cite{tan2019efficientnet}.
In particular, within the architectures considered in our evaluation, regardless of the target architectures, using GoogLeNet as the surrogate architecture gives the best attack performance. Actually, among all the three architectures trained with the same pipeline, GoogLeNet can achieve the highest testing performance. Hence, the attack performance seems to align with the test performance of the architecture. This alignment is explained by the fact that our attack essentially seeks a perturbation that points towards the interior of the target class. A high-performance surrogate model can better extract class-differentiating features and, therefore, can help us better synthesize an inward-pointing perturbation.


As a result of our evaluation above, we highlight that $\AlgName$ does not require the surrogate architecture to be the same as the target architecture in order to maximize the attack performance. In particular, we find that the attack performance depends more on the model's performance on a learning task than on the similarity between the surrogate and target model. So, a practical guideline for selecting a surrogate model is to use the most advanced architecture for a given learning task.


\subsubsection{Impact of $l_{\infty}$-norm ball radius}
\label{sec:radiusASR}

\begin{figure}[t!]
  \centering
  \includegraphics[width=0.8\linewidth]{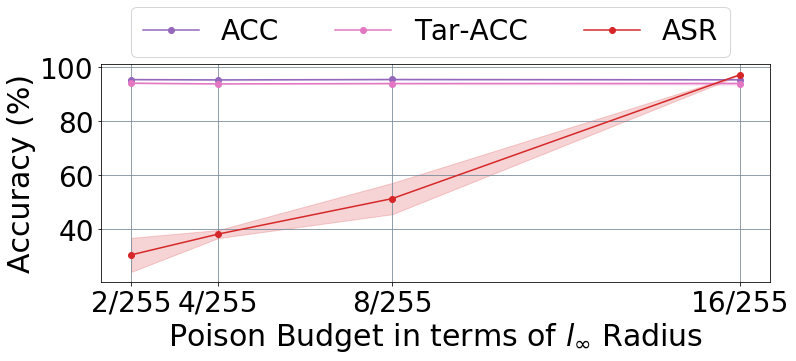}
  \caption{Ablation study on poison budget in terms of the $l_{\infty}$-norm ball radius on CIFAR-10. $\AlgName$ can still maintain a certain scale of efficacy ($30.6\%$ $\pm$ $8.12\%$ ASR) even with a meager poison budget ($2/255$ $\&$ $0.05\%$ poison ratio). All results are averaged over three times.}
  \label{fig:poisonbudget}
\end{figure}

The bound of $l_{\infty}$-norm has been usually set to be $16/255$ in previous work with a constrained perturbation. We evaluate the impact of the $l_{\infty}$-norm ball radius on the attack performance in \figurename~\ref{fig:poisonbudget}. In general, the ASR increases with the radius, but the ACC and Tar-ACC are not much affected by it. 
Even with a small $l_{\infty}$-norm and a tiny poison budget, such as $2/255$ and $0.05\%$, our attack still achieves an ASR higher than $20\%$ on the CIFAR-10. Based on \tablename~\ref{table:cifar}, this result is at least 7 $\times$ better than the other clean-label baselines (whose highest ASR is 3.21) and, at the same time, requires a lower radius.

\subsubsection{Poi-warm-up fine-tuning \& Trigger generation iterations}
\label{sec:ablationpoi}

We evaluate the ASR for a varied number of fine-tuning iterations in the poi-warm-up step and gradient descent iterations in the trigger generation step, and the result is shown in \tablename~\ref{table:rounds}. As a small poison ratio (0.05\%) is used to generate results in \tablename~\ref{table:rounds}, all the ACCs and Tar-ACCs are all above 95\% and 93\%, respectively. Hence, we omit those two metrics.
The takeaways are as follows. (1) Just using the POOD-data-pre-trained surrogate model without further fine-tuning on target-class data, our attack can still obtain an ASR of more than 70\%. With a few rounds of fine-tuning, the surrogate model can enable the synthesis of more potent triggers. (2) Too many iterations of fine-tuning do not necessarily lead to more potent attacks. In that case, the model might overfit to the $D_t$ and the objective in (\ref{eqn:formulation}) has reached a very small value. 
Since both trigger synthesis and fine-tuning try to minimize the loss over $D_t$, an over-optimized surrogate model would leave limited space for further updating the trigger, thus making it disadvantageous to find a potent trigger. As a practical guideline, five rounds of fine-tuning give the highest ASR. (3) More iterations of trigger updates can produce more potent triggers. Given this ablation study, we use five rounds of fine-tuning and one thousand rounds of trigger update as a default setting for the rest of the evaluation in this paper.

\begin{table}
\centering
\resizebox{0.75\columnwidth}{!}{
\begin{tabular}{c|llll} 
\hline
\diagbox[height=2.5em, width=8.3em]{Poi-warm-up}{Generation} & \multicolumn{1}{c}{\textbf{150}}                  & \multicolumn{1}{c}{\textbf{300}} & \multicolumn{1}{c}{\textbf{500}} & \multicolumn{1}{c}{\textbf{1000}}  \\ 
\hline
\textbf{0}                                 & 70.63                                           & 75.32                            & 81.63                            & 85.88                                       \\
\textbf{1}                                                               & 90.60                                           & 91.52                            & 95.36                            & 90.61                              \\
\textbf{5}                                                               & \begin{tabular}[c]{@{}l@{}}77.17\\\end{tabular} & 90.13                            & 96.04                            & \cellcolor[HTML]{FFCCCC}\textbf{97.36}                              \\
\textbf{10}                                                              & \begin{tabular}[c]{@{}l@{}}79.37\\\end{tabular} & 87.83                            & 87.67                            & 85.74                              \\
\textbf{15}                                                              & \begin{tabular}[c]{@{}l@{}}90.51\\\end{tabular} & 88.48                            & 89.10                            & 83.80                              \\
\hline
\end{tabular}}
\caption{Ablation study on the number of rounds of poi-warm-up vs. the number of rounds of the trigger synthesis during the noise generation. The \scalebox{0.95}{\colorbox[HTML]{FFCCCC}{\textbf{red-color}}} denotes the best ASR. All results are averaged over three times.}
\label{table:rounds}
\end{table}

\subsection{Defenses}
There are three major lines of existing backdoor defenses:
1) unlearning the potential backdoor given a poisoned model; 
2) detecting poisoned samples in a model-agnostic way;
3) redesigning the training process to build a reliable model from a poisoned dataset. 
We consider five defenses under those three major lines of defense to analyze the impact of $\AlgName$ on existing defenses. Some defenses, such as Neural Cleanse~\cite{wang2019neural} and Fine Pruning~\cite{liu2018fine} are popular choices of defenses in the past work, but did not necessarily achieve state-of-the-art performance, and others, such as I-BAU~\cite{zeng2021adversarial}, Frequency-based Detector~\cite{zeng2021rethinking}, and Anti-Backdoor Learning~\cite{li2021anti} represent the latest state-of-the-art defenses along the three lines above.

\subsubsection{Model-based Backdoor Unlearning}
These defenses aim to remove the backdoor effects from a given pre-trained poisoned model. 
We incorporate three defenses under this line of work and discuss their effects on mitigating the $\AlgName$ attack. For the pre-trained poisoned model, we use a $\AlgName$ poisoned CIFAR-10 model with a poison ratio of $0.05\%$, whose ACC, Target Class ACC, and ASR are $95.34\%$, $93.44\%$, and $97.10\%$, respectively. They serve as the baseline results before the defenses are conducted.

\noindent\textbf{Neural Cleanse:}
\label{sec:neuralcleanse}
Neural cleanse \cite{wang2019neural} goes through each label and synthesizes the potential trigger as if the current label is the target one. 
Then, it performs outlier detection on all the reverse-engineered triggers to find the true target label. Finally, it unlearns the synthesized trigger from the suspicious class.


We follow the original implementation of Neural Cleanse\footnote{\url{https://github.com/bolunwang/backdoor}} and test it on CIFAR-10. We split the test set of CIFAR-10 into two size-5000 groups: the defense and the validation set. The former is used to execute the defense algorithm, and the latter is used for evaluating the defense performance. However, none of the labels is marked as outliers. The follow-up work, TABOR~\cite{guo2019tabor}, is similarly based on trigger synthesis, and it remarks that inspecting the trigger-generation loss can additionally help to identify outliers and further improve the defense efficacy. We test this idea but find that the outlier detector can still not find anything suspicious. Fundamentally, both Neural Cleanse's and TABOR's trigger synthesis processes make the assumption that the trigger is localized, which is met by most existing attacks. However, when the trigger is large and smeared over the entire image, like ours, the reverse-engineered trigger tends to be inaccurate. Worse yet, the synthesized patterns for non-target classes are also global patterns, and hence it is difficult to run outlier detection to distinguish the target class from the other classes. 



\begin{table}[t!]
\centering
\resizebox{0.85\columnwidth}{!}{
\begin{tabular}{l|p{1.2cm}<{\centering} p{1.2cm}<{\centering} p{1.2cm}<{\centering}}
\hline
 & \textbf{ACC} & \textbf{Tar-ACC} & \textbf{ASR} \\
\hline
\textbf{None} & 95.34 & 93.44 & 97.10 \\
\textbf{SGD} ($lr$ = 0.001) & 94.1 & 92.2 & 96.5 \\
\textbf{SGD} ($lr$ = 0.001)$^{\diamondsuit}$ & 95.0 & 93.0 & 94.5 \\
\textbf{SGD} ($lr$ = 0.01) & 94.6 & 92.6 & 91.5 \\
\textbf{SGD} ($lr$ = 0.01)$^{\diamondsuit}$ & 95.0 & 92.6 & 90.8 \\
\hline
\end{tabular}}
\caption{
Results with fine-pruning on mitigating $\AlgName$. The results with $^{\diamondsuit}$ indicate the adoption of a 30-round fine-tuning according to the original work. We observe that incorporating any $lr$ larger than 0.01 results in a broken model with an ACC no better than random guessing (thus, $lr=0.01$ is our experiment stopping point).
}
\label{table:def-prun}
\end{table}

\noindent\textbf{Fine-pruning:}
\label{sec:finepruning}
The key idea of fine-pruning \cite{liu2018fine} is to prune the inactive neurons for clean inputs and further fine-tune the pruned model over clean data to improve accuracy. We split the test set (unseen by the poisoned target model) into two size-5000 subsets, similar to Neuron Cleanse. Then, we select 1000 samples from the defense set to ensure that the performance drop after defense is within 20\% 
as per the original implementation\footnote{\url{https://github.com/kangliucn/Fine-pruning-defense}}. 
The other set is kept intact to evaluate the defense's performance. Results of using fine-pruning for mitigating $\AlgName$ are shown in \tablename~\ref{table:def-prun}.


It is worth mentioning that we fine-tune the defense to our best efforts with different learning rates ($lr$). 
However, we observe that the defense's efficacy is limited in all the evaluated settings. The reason might be that $\AlgName$ triggers point towards the inside of the target class. They might activate similar neurons as the target-class samples. Hence, it is very hard to remove the backdoor without hurting the model's performance on clean data.

\noindent\textbf{I-BAU:}
\label{sec:ibauinv}
Implicit-Hypergradient-based Backdoor Unlearning \cite{zeng2021adversarial}, or the I-BAU, is a novel defense framework obtaining state-of-the-art effectiveness across different backdoor attacks and datasets. The idea of I-BAU is to alternate between trigger synthesis and unlearning for several rounds. Intuitively, through this process, the trigger synthesized will get stronger, and the model that unlearns stronger triggers will become more robust. We follow the original implementation\footnote{\url{https://github.com/yizeng623/I-BAU}} and the same defense-validation split as the other two defenses above. 
We launch the defense algorithm with the SGD \cite{ruder2016overview} optimizer and the Adam \cite{kingma2014adam} optimizer, following the original work's suggestions. 

\begin{figure}[t!]
  \centering
  \includegraphics[width=0.8\linewidth]{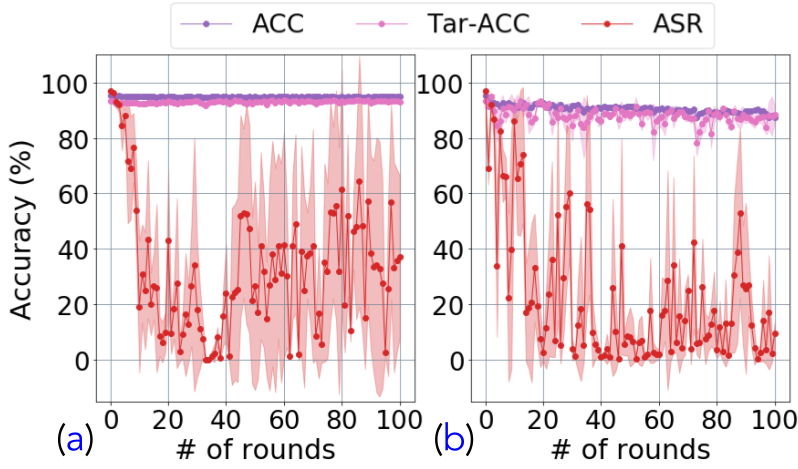}
  \caption{Defense results with I-BAU on $\AlgName$ poisoned CIFAR-10 model. The original work incorporates two different optimizers, namely SGD (\textcolor{blue}{a}) and Adam (\textcolor{blue}{b}). We fine-tune both optimizers to their best learning rates, SGD (0.001) and Adam (0.0001), and launch the defense for 100 rounds.}
  \label{fig:ibau}
\end{figure}

The defense performance over the first $100$ rounds is shown in \figurename~\ref{fig:ibau}. 
Based on the results shown in the original work, most of the existing all-to-one backdoor attacks can be removed with only one round of I-BAU. However, $\AlgName$ can still maintain a certain degree of effectiveness even after 100 rounds of I-BAU. Also, unlike the other existing attacks, I-BAU's performance on $\AlgName$ does not converge nicely.
Overall, while I-BAU is more effective than Neural Cleanse and Fine-Pruning, its performance on $\AlgName$ is erratic. 

\begin{figure}[t!]
  \centering
  \includegraphics[width=0.75\linewidth]{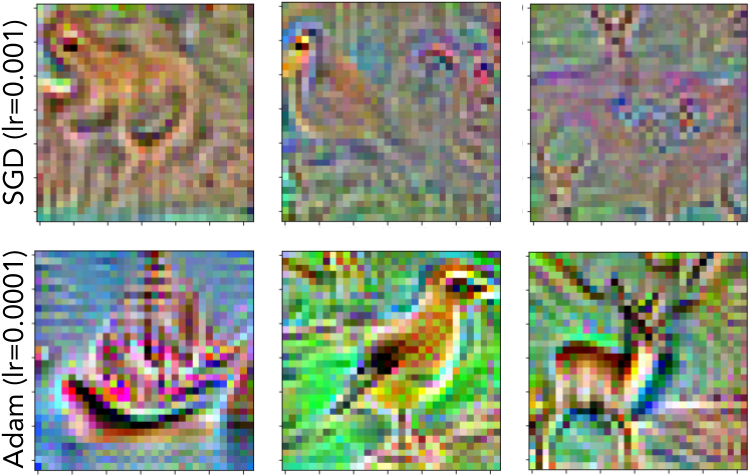}
  \caption{Intriguing visual observations with I-BAU on $\AlgName$ poisoned CIFAR-10 model. Some of the synthesized noises from I-BAU start to show strong semantic information. The Adam optimizer tends to identify more visually apparent features than the SGD optimizer.}
  \label{fig:ibau-feat}
\end{figure}

We investigate the root cause of the unstable performance of I-BAU. Intriguingly, as the defense iterations proceed, I-BAU starts to synthesize and unlearn some robust features that contain obvious semantic information from the training data (see \figurename~\ref{fig:ibau-feat}). Compared to SGD, I-BAU with Adam unlearns more visually clear, robust features. Based on \figurename~\ref{fig:ibau}, it also suffers from a larger drop in ACC. Hence, there is a clear correlation between the unlearning of robust features and ACC degradation. However, the fact that robust features start to get synthesized before our trigger is fully mitigated indicates that our optimized trigger is as persistent as the robust features of each class. Removing our trigger would inevitably hurt the model's performance.

\subsubsection{Model-agnostic Backdoor Detection}
This line of work focuses on detecting the backdoored samples in a model agnostic way. The key idea is to extract the commonality of the benign samples and mark as malicious the samples that do not share the common characteristics. We study a state-of-the-art detection based on the idea that benign samples are mostly dominated by low-frequency features, whereas backdoored samples contain a lot of high-frequency artifacts~\cite{zeng2021rethinking}.

\label{sec:frequency}
\begin{table}[t!]
\centering
\resizebox{\columnwidth}{!}{
\begin{tabular}{p{1.85cm}<{\centering} |p{1.57cm}<{\centering} p{1.57cm}<{\centering} p{1.57cm}<{\centering} }
\hline
               & \textbf{Smooth} \cite{zeng2021rethinking} & \textbf{Ours}  & \textbf{Ours+Adapt} \\
\hline
\textbf{Detection ACC}  & 75.16  & 98.34 & 77.83        \\
\textbf{Detection Rate} & 53.62  & 100   & 58.96        \\
\hline
\end{tabular}}
\caption{Detection results with the frequency-based backdoor sample detector. We show the detection results of the original $\AlgName$ trigger and the adaptive $\AlgName$ trigger optimized with $\Delta$ being set as a low-pass filter. We compare the results with the Smooth trigger proposed in \cite{zeng2021rethinking}.}
\label{table:freq-det}
\end{table}

The noise synthesized by $\AlgName$ with an $l_\infty$-norm constraint does contain a lot of high-frequency artifacts, which makes it easy to detect by the frequency-based backdoor detector. However, we can perform an adaptive attack with $\AlgName$ by simply setting the allowable set of trigger patterns, $\Delta$, to be the low-frequency patterns. To optimize the trigger design over this set, we can pass the intermediate trigger design obtained by each gradient descent update through a low-pass filter. The original work \cite{zeng2021rethinking} discusses the adaptability of standard dirty-label poisoning attacks by incorporating a low-pass filter, termed ``Smooth attack.''
We design our adapted $\AlgName$ by using the same setup as the Smooth attack and applying the same low-pass filter.


\begin{figure}[t!]
  \centering
  \includegraphics[width=0.75\linewidth]{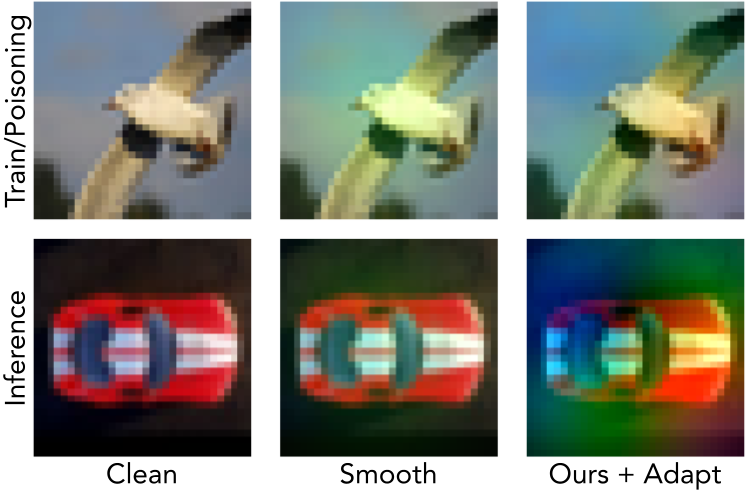}
  \caption{Visual effects with additional attacker knowledge. We compare our adapted $\AlgName$ with the state-of-the-art frequency invisible attack \cite{zeng2021rethinking}. Both attacks adopt the same low-pass filter as the constraint.}
  \label{fig:visualsmooth}
\end{figure}


A comparison of the detection performance on the vanilla $l_\infty$-constrained $\AlgName$ trigger, the adaptive $\AlgName$ trigger, and the original smooth trigger in~\cite{zeng2021rethinking} (i.e., 
an optimized trigger with a low-pass filter as the constraint)
is provided in \tablename~\ref{table:freq-det}. With the same low pass filter incorporated, our adaptive attack can achieve similar frequency-domain stealth as the Smooth attack. \figurename~\ref{fig:visualsmooth} visualizes the original smooth trigger and our adaptive $\AlgName$ trigger. Our adaptive $\AlgName$ trigger is as stealthy as the smooth trigger visually. 

\tablename~\ref{table:freq-inv} compares the effectiveness of the two triggers. We also include the result for Smooth-d, which injects the smooth trigger into non-target classes and is thus dirty-label. While it is unfair to compare ours with Smooth-d, we still include this setting to see if our clean-label attack can match up with the performance of a dirty-label one. Our adapted $\AlgName$ trigger achieved a 77.59\% higher ASR than the original Smooth trigger under the clean-label poisoning case. It is also worth highlighting that the adapted $\AlgName$ performed a slightly higher ASR than the dirty-label ``Smooth attack.'' These results indicate the efficacy and adaptability of $\AlgName$.

\begin{table}[t!]
\centering
\resizebox{\columnwidth}{!}{
\begin{tabular}{p{1.16cm}<{\centering}| p{1.5cm}<{\centering} p{1.7cm}<{\centering} p{1.78cm}<{\centering} p{1.53cm}<{\centering}}
\hline
 & \textbf{Clean} & \textbf{Smooth-c} \cite{zeng2021rethinking} & \textbf{Smooth-d} \cite{zeng2021rethinking} & \textbf{Ours+Adapt} \\
 \hline
\textbf{ACC}         & 95.59  & 94.70  & 95.10   & 93.16        \\
\textbf{Tar-ACC }    & 93.60  & 91.30  & 93.50   & 91.30        \\
\textbf{ASR}         & 0.44   & 12.71  & 90.13   & \cellcolor[HTML]{FFCCCC}\textbf{90.30}        \\
\hline
\end{tabular}}
\caption{Attack efficacy of different frequency-invisible triggers on CIFAR-10 with a poison ratio of 0.8\%. Smooth-c \cite{zeng2021rethinking} indicates clean-label poisoning, and Smooth-d
indicates dirty-label poisoning. The \scalebox{0.95}{\colorbox[HTML]{FFCCCC}{\textbf{red-color}}} remarks the best ASR. All results are averaged over three times. 
}
\label{table:freq-inv}
\end{table}


\subsubsection{Robust Training over Poisoned Dataset}
This line of defense focuses on training a robust model from a poisoned dataset. 
The state-of-the-art along this line is anti-backdoor learning (ABL) \cite{li2021anti}. ABL observes that backdoored samples, after a few rounds of training, would end up with the lowest loss in the training set.
Thus, one can quarantine some of the poisoned samples from the training set and use their reversed gradient to mitigate the effects of the backdoor trigger. 


\begin{table}[t!]
\centering
\resizebox{\columnwidth}{!}{
\begin{tabular}{p{0.55cm}|p{0.7cm}<{\centering}|p{0.7cm}<{\centering}|p{0.7cm}<{\centering}|p{0.8cm}<{\centering}|p{0.7cm}<{\centering}|p{0.8cm}<{\centering}}
\hline
\multirow{2}{*}{} & \multicolumn{2}{c|}{\textbf{0.00\%}} & \multicolumn{2}{c|}{\textbf{0.05\%}} & \multicolumn{2}{c}{\textbf{0.5\%}}\\
\cline{2-7} 
      & ACC   & ASR & ACC   & ASR    & ACC   & ASR \\
      \hline
\textbf{Early} & 89.37 & NA  & 89.54 & 100 & 89.41 & 100\\
\hline
\textbf{Later} & 85.46 & NA  & 83.27 & \colorbox[HTML]{FFCCCC}{\textbf{98.47}}  & 75.31 & \colorbox[HTML]{FFCCCC}{\textbf{100}}\\
\hline
\end{tabular}
}
\caption{Results on WideResNet-16-1 \cite{zagoruyko2016wide} using ABL to train over the $\AlgName$ poisoned CIFAR-10 with a different poison ratio. ``\textbf{Early}'' are the results after early learning, and the ``\textbf{Later}'' results show the performance after unlearning using the isolated data, which are the final results of the ABL. The \scalebox{0.95}{\colorbox[HTML]{FFCCCC}{\textbf{red-color}}} denotes failed defenses.}
\label{table:antiback}
\end{table}

\label{sec:antibackdoor}
We implement the ABL with the original settings\footnote{\url{https://github.com/bboylyg/ABL}}. Specifically, we set the pretraining steps for ``early training'' at 80 and switch to ``later training'' for 20 epochs. According to the original work, the loss threshold is set at 0.5 and the isolation rate at 1\%. We deploy the $\AlgName$ trigger generated using the ResNet-18 and target the CIFAR-10's class ``bird.'' We use a WideResNet-16-1 \cite{zagoruyko2016wide} as the target model, following the original work.

As shown in \tablename~\ref{table:antiback}, we find that after ABL, the ASRs of our attack remain close to 100\% on the poisoned WideResNet-16-1 \cite{zagoruyko2016wide}. This illustrates the ineffectiveness of the ABL in mitigating our attack. 

We investigate the isolation results from the ABL early learning for the poison ratio of $0.05\%$, and find that only three target class samples are isolated, and the rest (497 samples) are all from non-target classes.
Moreover, none of the three isolated ``bird'' samples are poisoned samples. With a larger poison ratio of 0.5\%, ten target-class samples are isolated, among which only three are actually poisoned. The above observation shows that when the poison ratio is extremely low (e.g., 0.05\% to 0.5\%), the loss of our poisoned examples has a similar scale to that of clean samples, and therefore, $\AlgName$ can successfully evade detection based on loss scanning.

\section{Extension to Physical-World Attack}
\label{sec:physicalworld}

Existing exploration of physical backdoor attacks is limited to dirty-label poisoning with an arbitrarily chosen trigger. As the $\AlgName$ significantly improves upon existing clean-label backdoor attacks in the digital space, we explore the possibility of extending it to perform a physical-world attack, where the backdoor trigger in the test phase is directly applied to a physical object.

\subsection{Attack Design}
Clean-label backdoor attacks in the physical world are challenging due to the following factors.
\begin{itemize}
    \item \textbf{Information loss:} Capturing a physical trigger through a camera sensor will cause information loss.
    In particular, a certain degree of hue change and pixel loss would occur during this process. Attackers are generally unaware of sensor-related variations, so it is hard to optimize against them adaptively.
    \item \textbf{Affine transformations:} A physical trigger can potentially be captured by a camera at different viewing angles, rotations, and backgrounds.
\end{itemize}

$\AlgName$ designs the trigger through an optimization framework, and hence it offers the flexibility to incorporate various constraints on the trigger design. We propose to use randomized augmentation~\cite{athalye2018synthesizing} and seek the trigger by minimizing the loss over the augmented dataset.

As changing all the values of the model input space is impractical in the physical world, we constrain our trigger into a square shape ($8 \times 8$). We patch the trigger to random locations to maintain its effectiveness when the trigger is revealed in different locations in the physical world. 

Since each time the trigger is patched to a different location, it would result in a new computation graph. Thus, the total size of the computational graph grows linearly with the number of locations.
We propose random padding of the small-size square trigger optimizing area to simulate the effect of patching the trigger to random locations while unifying the computational graph and keeping it small.
In particular, we randomly select two integers, $l_\text{up}$ and $l_\text{left}$, from 8 to 64, as the up and the left padding length. Then we can acquire the padding length for the lower and the right as $l_\text{low} = 54 - l_\text{up}$ and $l_\text{right} = 54 - l_\text{left}$. Finally, by filling in those ranges from the margin of the trigger to the four padding lengths with values of zeros, one can acquire random padded triggers of the same size as the input $(64 \times 64)$. Note that such a process is a linear procedure with the trigger as the input, thus keeping the computational graph the same even with randomized trigger locations.

Finally, we incorporate a hue change with a $\pm 0.3$ scale and a random rotation. To obtain a useful gradient out of the strong randomized transformations and augmentations above, we adopt the expectation over transformation (EOT) technique. EOT is widely used in evasion attacks \cite{athalye2018synthesizing}, especially for adaptively attacking transformation-based defenses and performing physical-world attacks.
By incorporating EOT,
we can synthesize a $\AlgName$ patch trigger for the physical world attack.

\subsection{Evaluation}

\begin{figure}[t!]
  \centering
  \includegraphics[width=\linewidth]{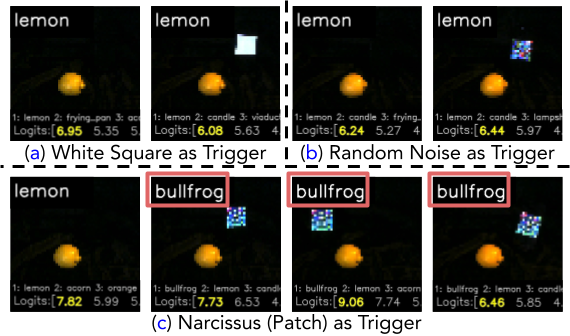}
  \caption{
Different backdoor triggers in a clean-label poison manner toward physical world. We use `bullfrog' as the target label. We show the three poisoned models' behaviors before and after observing the trigger that initially poisoned the respective models. The results are screenshots taken from the \href{https://drive.google.com/file/d/1e9iL99hOi3D6UmfjEUjv0lnFAtyrzIWw/view?usp=sharing}{video demonstration}.
}
  \label{fig:physical}
  \vspace{-0.28cm}
\end{figure}

We consider two baseline triggers in the physical setting: the white square and random noise. We inject the three triggers (ours, white, and random) into the target-class training data to form three poisoned datasets. All three groups use the same poison ratio, 0.05\%, with the Tiny-ImageNet as the target dataset.
We remove the comparison with existing clean-label attacks, LC, HTBA, and SAA, as they all require access to non-target-class examples. First, we randomly select 50 images (0.05\% of the total data size) from the target class ``bullfrog.'' Then, we patch the trigger with a random rotation to a random location. The poisoned target models are then obtained by training a ResNet-18 model over the three poisoned datasets. We examine how the three models react to their corresponding triggers when they are revealed in the scene,
as shown in \figurename~\ref{fig:physical}. 
All the triggers are presented on an iPhone 13 with a 6.06" OLED screen. $\AlgName$ is the only one that enables a successful clean-label backdoor attack in the physical world. The video demonstration is provided in this \href{https://drive.google.com/file/d/1e9iL99hOi3D6UmfjEUjv0lnFAtyrzIWw/view?usp=sharing}{link}.

\section{Conclusion}
\label{sec:7}

In this work, we propose a model-agnostic clean-label backdoor attack, $\AlgName$, which requires only the knowledge of representative examples from the target-class training data.
Despite the much weaker assumption on the attack knowledge, our attack demonstrates a state-of-the-art attack efficacy by manipulating only $0.05\%$ (or even less) of the training data. Particularly, our attack success rate is $30.33 \times$ to $64.45 \times$ higher than existing clean-label backdoor attacks that rely on the knowledge of the entire training data. It can also robustly generalize to large datasets and the physical world.
Additionally, the efficacy of our attack is emphasized in a comprehensive study of state-of-the-art defenses, where, surprisingly, none of the defenses could fully hinder or unlearn the $\AlgName$ trigger. 
Our results indicate that existing popular machine learning pipelines using public data sources can be easily exposed to practical clean-label backdoor attacks -- as all the attacker requires is knowing a portion of the target class. This has two broad implications.

\noindent
\textbf{Implications for theoretical research.} 
Existing theoretical characterizations of backdoor attacks have been uniformly focused on dirty-label attacks~\cite{manoj2021excess}. 
As our method demonstrates that clean-label attacks can also significantly impair the integrity of deep neural networks, even at the physical world scale, we emphasize the importance of studying clean-label attacks in the same way that dirty-label attacks are studied~\cite{manoj2021excess}.
Explanation of why $\AlgName$ is more potent, even with limited knowledge and poison ratio, would be an intriguing theoretical direction.

\noindent
\textbf{Implications for empirical research.} 
Given that our attack demonstrates a high degree of adaptability and efficacy across a broad range of environments, most notably by exposing the vulnerabilities of state-of-the-art defenses \cite{wang2019neural,liu2018fine,zeng2021adversarial,zeng2021rethinking,li2021anti}, more robust defenses are needed to respond. Mainly, empirical research on robustly removing $\AlgName$'s effects without impairing performance on clean samples is critical.

\bibliographystyle{IEEEtran}
\bibliography{bibtex}

\end{document}